\documentclass[aps,prl,twocolumn,showpacs,10pt,superscriptaddress,preprintnumbers]{revtex4-1}
\usepackage{amsmath}
\usepackage{graphicx}
\usepackage{amssymb}
\usepackage[colorlinks,citecolor=red]{hyperref}

\usepackage{color}

\newcommand{\beq}{\begin{equation}}
\newcommand{\eeq}{\end{equation}}
\newcommand{\bea}{\begin{eqnarray}}
\newcommand{\eea}{\end{eqnarray}}
\newcommand{\nn}{\nonumber}

\begin{document}

\preprint{
{\vbox {
\hbox{\bf MSUHEP-18-001}
}}}
\vspace*{0.2cm}

\title{Transverse Momentum Resummation for $t$-channel\\ single top quark production at the LHC}

\author{Qing-Hong Cao}
\email{qinghongcao@pku.edu.cn}
\affiliation{Department of Physics and State Key Laboratory of Nuclear Physics and Technology, Peking University, Beijing 100871, China}
\affiliation{Collaborative Innovation Center of Quantum Matter, Beijing 100871, China}
\affiliation{Center for High Energy Physics, Peking University, Beijing 100871, China}

\author{Peng Sun}
\email{pengsun@msu.edu}
\affiliation{Department of Physics and Astronomy,
Michigan State University, East Lansing, MI 48824, USA}
\affiliation{Department of Physics and Institute of Theoretical Physics, Nanjing Normal University, Nanjing, Jiangsu, 210023, China}

\author{Bin Yan}
\email{yanbin1@msu.edu}
\affiliation{Department of Physics and State Key Laboratory of Nuclear Physics and Technology, Peking University, Beijing 100871, China}
\affiliation{Department of Physics and Astronomy,
Michigan State University, East Lansing, MI 48824, USA}

\author{C.-P. Yuan}
\email{yuan@pa.msu.edu}
\affiliation{Department of Physics and Astronomy,
Michigan State University, East Lansing, MI 48824, USA}

\author{Feng Yuan}
\email{fyuan@lbl.gov}
\affiliation{Nuclear Science Division, Lawrence Berkeley National
Laboratory, Berkeley, CA 94720, USA}
\begin{abstract}

We study the effect of multiple soft gluon radiation on the kinematical distributions of the $t$-channel single top quark production at the LHC. By applying the transverse momentum dependent factorization formalism, large logarithms (of the ratio of large invariant mass $Q$ and small total transverse momentum $q_\perp$  of the single-top plus one-jet final state system) are resummed to all orders in the expansion of the strong interaction coupling at the accuracy of next-to-leading logarithm, including the complete next-to-leading order corrections. We show that the main difference from PYTHIA prediction lies on the inclusion of the exact color coherence effect between the initial and final states in our resummation calculation, which becomes more important when the final state jet is required to be in the forward region. We further propose to apply the experimental observable $\phi^*$, similar to the one used in analyzing precision Drell-Yan data, to test the effect of multiple gluon radiation in the single-top events. The effect of bottom quark mass is also discussed.

\end{abstract}

\maketitle

\noindent {\bf Introduction:~}
The top quark is the heaviest particle of the Standard Model (SM) of
elementary particle physics, 
with its mass around the electroweak symmetry breaking scale.
It is believed that studying its detailed interactions could shed light on possible
New Physics beyond the SM. Furthermore, the lifetime of the top quark is
much smaller than the typical hadronization time scale,
so that one can also determine the properties
(including polarization) of this heavy
bare quark, produced from various scattering processes, by studying the kinematical
distributions of the top quark and its decay particles.
Top quarks are predominantly produced in pairs through gluon fusion process, via strong interaction, at
the CERN Large Hadron Collider (LHC).
It can also be produced singly via charged-current electroweak interaction,
involving a $Wtb$ coupling~\cite{Dawson:1984gx,Willenbrock:1986cr,Dawson:1986tc,Yuan:1989tc},
which offers a promising way to precisely study the $Wtb$ coupling
and the $V_{tb}$ Cabibbo-Kobayashi-Maskawa (CKM) matrix element.

To test the $Wtb$ coupling of the top quark from
measuring the production rate of single-top events, one has to be able to
precisely predict the detection efficiency of the events after imposing 
needed kinematic cuts. Hence, higher order calculations are required.
The single top quark production and decay in hadron collision
at the next-to-leading order (NLO) and next-to-next-to-leading order (NNLO) accuracy in QCD correction have been discussed widely. (See Ref.~\cite{Berger:2017zof}, and the references therein).  
To go beyond the fixed-order calculations,  the threshold resummation technique has been applied to improve the prediction on the single-top inclusive production rate at the next-to-leading-logarithm (NLL) and next-to-next-to-leading-logarithm (NNLL) accuracy~\cite{Kidonakis:2006bu,Kidonakis:2007ej,Zhu:2010mr,Wang:2010ue,Kidonakis:2011wy,Wang:2012dc}.  The threshold resummation technique has also been used to improved the prediction on the transverse momentum distribution of the top quark by summing over large logarithms $\ln(m_t^2/s_4)$ with $s_4\to 0$, where $s_4=\hat{s}+\hat{t}+\hat{u}-m_t^2$, $\hat{s}$, $\hat{t}$ and $\hat{u}$  are the usual Mandelstam variables~\cite{Kidonakis:2006bu,Kidonakis:2007ej,Zhu:2010mr,Wang:2010ue,Kidonakis:2011wy,Wang:2012dc}.

In this Letter, we focus on improving the prediction on
the kinematical distributions of $t$-channel single top events,
by applying the transverse momentum resummation formalism to sum over
large logarithms $\ln(Q^2/q_{\perp}^2)$, with $Q \gg q_{\perp}$,
 to all orders in the expansion of the strong interaction coupling at
the NLO-NLL accuracy, where $Q$ and $q_\perp$ are the
invariant mass $Q$ and total transverse momentum
$q_\perp$  of the single-top plus one-jet final state system, respectively.
We adopt the $q_\perp$ resummation formalism based on the transverse momentum
dependent (TMD) factorization formalism ~\cite{Collins:1984kg}, which has
been widely discussed in the literature to resum this sort of
large logarithms in the color singlet processes, such as the
Drell-Yan pair production~\cite{Collins:1981uk,Collins:1981va}.
The application of the $q_\perp$ resummation formalism
for processes with more complicated color structures, such as heavy quark
production,  was firstly discussed in Refs.~\cite{Zhu:2012ts,Li:2013mia,Zhu:2013yxa}.
For processes involving massless jets in the final state,
the $q_\perp$ resummation formalism needs to be further modified to take into account
the color coherence effect induced by the presence of the light (quark and gluon) 
jets in the final state~\cite{Sun:2014lna,Sun:2014gfa,Sun:2015doa,Sun:2016mas,Sun:2016kkh,Xiao:2018esv}.
The extra soft gluon radiations in the event could be either 
within or outside the observed final-state jet cone.
Within the jet cone, the radiated gluon is treated as collinear to the final state parton, and it leads to a
contribution to the bin of $q_{\perp}=0$. This contribution can be factorized out as a jet function based on
the TMD resummation formalism~\cite{Sun:2015doa}.
When outside of observed final-state jet cone, the radiated soft gluon will generate a non-vanishing $q_\perp$,
and induce the large logarithms $\ln(Q^2/q_{\perp}^2)$ which needs to be resummed via the {\it modified} $q_\perp$ resummation formalism.

The experimental signature of the $t$-channel single top event at the LHC is  
an energetic light jet, associatively produced with the 
single top quark, in the final state.  
As to be shown below, the location and height of the
Sudakov peak, in the $q_\perp$ distribution of $t$-channel single top events,
strongly depends on the color coherence effect, induced by 
soft gluon interaction between the initial and final state jets,
and the treatment of bottom quark mass in the resummation calculation.
The (formally) sub-leading logarithms play an important role
when the final state jet is required to be in the forward region, 
where our resummation prediction is noticeably different from 
the PYTHIA parton shower result.

\noindent {\bf Resummation Formalism:~}%
We consider the process $pp\to t +jet+X$ at the LHC.
Using the TMD resummation formalism  presented
in Ref.~\cite{Sun:2015doa}, the differential cross section of the
$t$-channel single top quark production process
can be summarized as
\begin{align}
&\frac{d^4\sigma}
{dy_t dy_J d P_{J\perp}^2
d^2q_{\perp}}=\sum_{ab}\nn\\
&\left[\int\frac{d^2\vec{b}}{(2\pi)^2}
e^{-i\vec{q}_\perp\cdot
\vec{b}}W_{ab\to t J}(x_1,x_2,\textbf{b})+Y_{ab\to tJ}\right] \ ,\label{resumy}
\end{align}
where $y_t$ and $y_J$ are the rapidities for the top quark and the final state jet, respectively;
 $P_{J\perp}$ and $q_{\perp}$ are the transverse momenta of the jet and the total transverse momentum of the top quark and the jet system, i.e. $\vec{q}_\perp=\vec{P}_{t\perp}+\vec{P}_{J\perp}$. The $W_{ab\to tJ}$ term contains all order
 resummation contribution, in powers of $\ln(Q^2/q_{\perp}^2)$, and the inclusion of
 the $Y_{ab\to tJ}$ term is to account for the missing (non-singular) 
 part of fixed-order correction when expanding 
  the $W_{ab\to tJ}$ term to the same order in the strong coupling constant $g_s$ 
  as the fixed order calculation. 
 The variables $x_1, x_2$ are momentum fractions of the incoming hadrons carried by 
 the two incoming partons.

The above $W$ term can be further written as
\begin{widetext}
\begin{eqnarray}
W_{ab\to tJ}\left(x_1,x_2,\textbf{b}\right)&=&x_1\,f_a(x_1,\mu_F=b_0/b_*)
x_2\, f_b(x_2,\mu_F=b_0/b_*) e^{-S_{\rm Sud}(Q^2, \mu_{\rm Res},b_*)}e^{-\mathcal{F}_{NP}(Q^2,\textbf{b})} \nonumber\\
&\times& \textmd{Tr}\left[\mathbf{H}_{ab\to tJ}(\mu_{\rm Res})
\mathrm{exp}[-\int_{b_0/b_*}^{\mu_{\rm Res}}\frac{d
\mu}{\mu}\mathbf{\gamma}_{}^{s\dag}]\mathbf{S}_{ab\to tJ}(b_0/b_*)
\mathrm{exp}[-\int_{b_0/b_*}^{\mu_{\rm Res}}\frac{d
\mu}{\mu}\mathbf{\gamma}_{}^{s}]\right]\ ,\label{resum}
\end{eqnarray}
\end{widetext}
where $Q^2=\hat{s}=x_1x_2S$, $b_0=2e^{-\gamma_E}$,
$f_{a,b}(x,\mu_F)$ are parton distribution functions (PDF) for the incoming partons $a$ and $b$,
and $\mu_{\rm Res}$ represents the resummation scale of this process.
Here, we define $b_*=\textbf{b}/\sqrt{1+\textbf{b}^2/b_{\rm{max}}^2}$ with $b_{\rm {max}}=1.5~{\rm GeV}^{-1}$,
which is introduced to factor out the non-perturbative contribution
$e^{-\mathcal{F}_{NP}(Q^2,b)}$, arising from the large $\textbf{b}$ region (with $\textbf{b} \gg b_*$)
~\cite{Landry:1999an,Landry:2002ix,Sun:2012vc}.
In this study, we shall use CT14NNLO PDFs~\cite{Dulat:2015mca} for our numerical calculation.
Hence, our resummation calculation should be consistently done in the
General-Mass-Variable-Flavor (GMVR) scheme in which the PDFs are
determined.
The bottom quark PDF is set to zero when the factorization scale $\mu_F$
is below the bottom quark mass $m_b$.
To properly describe the small $q_\perp$ region (for $q_\perp<m_b$), the S-ACOT scheme~\cite{Aivazis:1993kh,Aivazis:1993pi,Collins:1998rz,Kramer:2000hn} is
adopted to account for the effect from the (non-zero) mass of the incoming bottom quark
in the hard scattering process $q b \to q^{\prime}t + X$.
In Refs.~\cite{Nadolsky:2002jr,Belyaev:2005bs,Berge:2005rv}, a detailed discussion has been given on how to implement the
S-ACOT scheme in the $q_\perp$ resummation formalism, for processes initiated by
bottom quark scattering. 
In short,
the S-ACOT scheme retains massless quark in the calculation of
the hard scattering amplitude (of $q b \to q^{\prime}t$),
 but with the (bottom quark) mass dependent Wilson coefficient $C_{b/g}^{(1)}(x,\textbf{b},\mu_F)$,
 to account for the contribution from gluon splitting into a $b\bar{b}$ pair~\cite{Nadolsky:2002jr,Belyaev:2005bs,Berge:2005rv}.
The hard and soft factors $\mathbf{H}$ and $\mathbf{S}$ are expressed as matrices in the color space of  $ab\to tJ$, and $\gamma^s$ is the associated anomalous dimension for the soft factor. The Sudakov form factor ${\cal S}_{\rm Sud}$ resums the leading double logarithm and the sub-leading logarithms, and is found to be 
\begin{align}
S_{\rm Sud}(Q^2,\mu_{\rm Res},b_*)=\int^{\mu_{\rm Res}^2}_{b_0^2/b_*^2}\frac{d\mu^2}{\mu^2}
\left[\ln\left(\frac{Q^2}{\mu^2}\right)A+B \right.\nn\\
\left.+D_1\ln\frac{Q^2-m_t^2}{P_{J\perp}^2R^2}+
D_2\ln\frac{Q^2-m_t^2}{m_t^2}\right]\ , \label{su}
\end{align}
where $R$ represents the cone size of the final state jet, $m_t$ is the top quark mass. Here, the parameters $A$, $B$, $D_1$ and $D_2$ can be expanded perturbatively in $\alpha_s$, which is $g_s^2/(4 \pi)$. At one-loop order,
\begin{equation}
A=C_F\dfrac{\alpha_s}{\pi},\quad B=-2C_F\dfrac{\alpha_s}{\pi},\quad D_1=D_2=C_F\dfrac{\alpha_s}{2\pi},
\end{equation}
with $C_F=4/3$.
In our numerical calculation, we will also include the $A^{(2)}$ contribution since it is associated with the incoming parton distributions and universal for all processes initiated by the same incoming partons. The cone size $R$ is introduced to regulate the collinear gluon radiation associated with the final state jet~\cite{Sun:2014lna,Sun:2014gfa,Sun:2015doa,Sun:2016mas,Sun:2016kkh,Xiao:2018esv}.

The soft gluon radiation can be factorized out based on the Eikonal approximation method. 
For each incoming or outgoing color particle, the soft gluon radiation is factorized into an associated gauge link along the particle momentum direction. The color correlation between the color particles in this process can be described by a group of orthogonal color bases.
For the $t$-channel single top quark production, there are two orthogonal color configurations, 
which are 
\begin{equation}
C_{1kl}^{ij}=\delta_{ik}\delta_{jl},\quad C_{2kl}^{ij}=T^{a^\prime}_{ik}T^{a^\prime}_{jl},
\end{equation}
where $i,j$ are color indices of the two incoming partons, $k,l$ are color indices of the jet and the top quark in final states and $a^\prime$ is color index of the gluon. We follow the procedure of Ref.~\cite{Sun:2015doa} to calculate the soft factor. Its definition in such color basis can be written as
\begin{align}
S_{IJ}&=\int_0^\pi
\frac{d\phi}{\pi}\; C^{bb'}_{Iii'} C^{aa'}_{Jll'}\langle 0|{\cal L}_{
vcb'}^\dagger(b) {\cal
L}_{\bar vbc'} (b){\cal L}_{\bar
vc'a'}^\dagger(0)\nn\\
&\times {\cal
L}_{ vac}(0) {\cal L}_{n ji}^\dagger(b) {\cal
L}_{\bar n i'k}(b) {\cal L}_{\bar nkl}^\dagger (0) {\cal
L}_{nl'j} (0)  |0\rangle \ ,\label{soft}
\end{align}
where we integrated out the azimuthal angle of the top quark and traded the relative azimuthal angle $\phi$ for  the $q_\perp$. $I$ and $J$ represent the color basis index, $n$ and $\bar{n}$ represent  the momentum directions of the top quark and the jet in this process,  $v$ and $\bar{v}$ are the momentum directions of the initial states.

The anomalous dimension of the soft factor $S_{IJ}$ can be calculated at one-loop order 
and found to be
\begin{align}
\mathbf{\gamma^S}_{ub\to dt}&=\dfrac{\alpha_s}{\pi}\left[\begin{array}{cc}
      C_F\,T &\;\; C_F/C_A\,U \\\\
       U &\;\; \frac{1}{2}(C_A-2/C_A)\,U-\frac{1}{2C_A}\,T \\
     \end{array} \right ]\ \ , 
\label{eq:gammas}
\end{align}
where,
\begin{align}
T&=\ln(\dfrac{-\hat{t}}{\hat{s}})+\ln(\dfrac{-(\hat{t}-m_t^2)}{\hat{s}-m_t^2}),&\\
U&=\ln(\dfrac{-\hat{u}}{\hat{s}})+\ln(\dfrac{-(\hat{u}-m_t^2)}{\hat{s}-m_t^2}).
\label{sud15}
\end{align}
Here $C_A=3$, $\hat{t}=(p_u-p_d)^2$, $\hat{u}=(p_b-p_d)^2$ for the $ub\to dt$ process.

The hard factor $H_{IJ}$ contains the contribution from the jet function which is proportional to the leading order cross section. 
The jet function accounts for contribution originated from collinear gluon radiation, and is  dependent on the jet algorithm used in the calculation. In this work, we apply the anti-$k_T$ jet algorithm, as discussed in Refs.~\cite{Mukherjee:2012uz,Sun:2015doa}.

Before concluding this section, we would like to point it out that we did not include in this work the possibility of non-global logarithms~\cite{Dasgupta:2001sh,Dasgupta:2002bw,Banfi:2003jj,Forshaw:2006fk}.
The non-global logarithms (NGLs) arise from some special kinematics of two soft gluon radiations, in which the first one is radiated outside of the jet which subsequently radiates a second gluon into the jet. 
We have roughly estimated its numerical effect and found that 
the NGLs are negligible in this process since it starts at $\mathcal{O}(\alpha_s^2)$~\cite{Sun:temp}. 
Therefore, we will ignore their contributions in the following phenomenology discussion.

\begin{figure}
\includegraphics[width=0.234\textwidth]{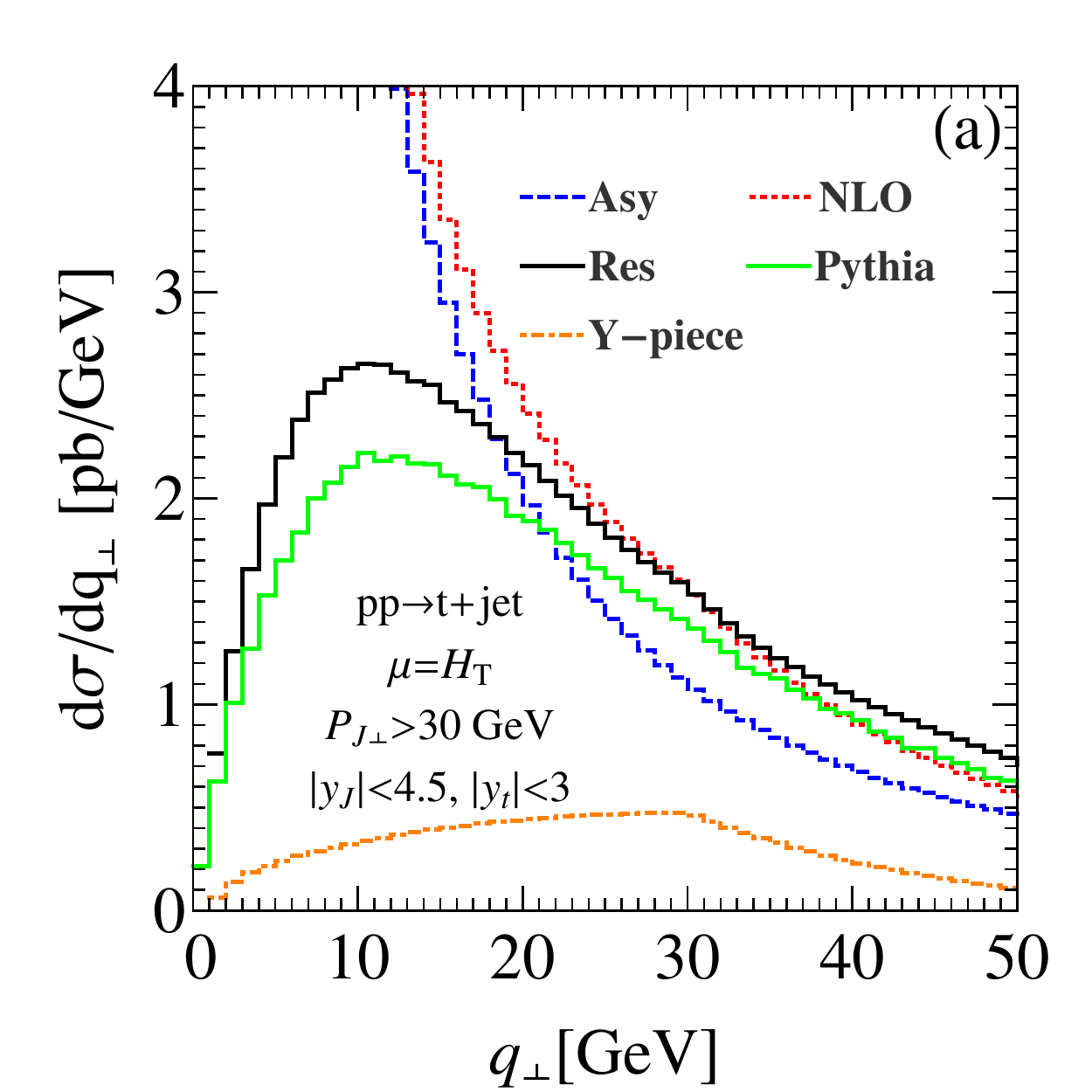}
\includegraphics[width=0.234\textwidth]{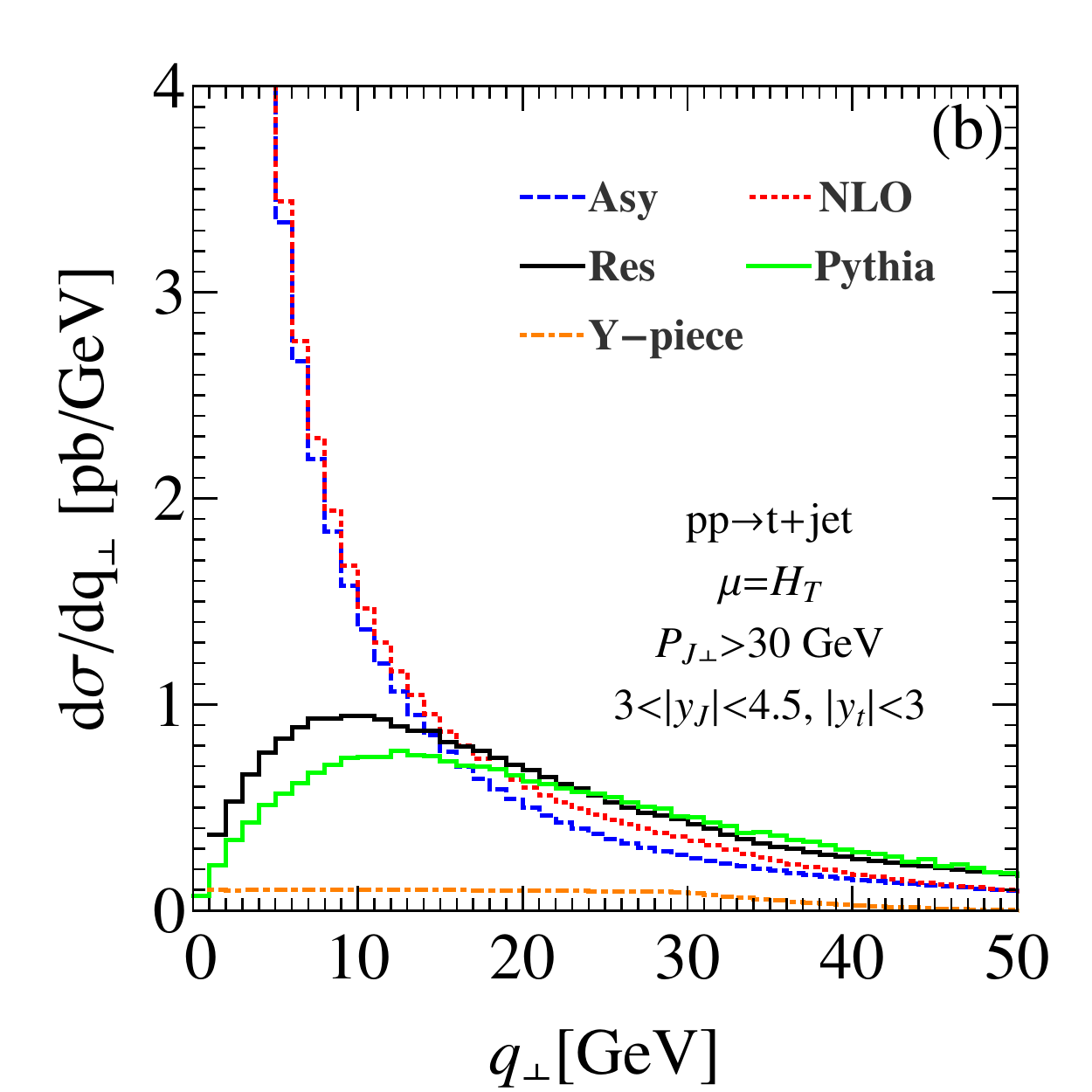}
\caption{ The $q_{\perp}$ distribution from the asymptotic result (blue dashed line), NLO calculation (red dotted line), resummation prediction (black solid line), parton shower result by PYTHIA 8 (green solid  line) and $Y$-term (orange dot-dashed  line) for the $t$-channel single top quark production  at the $\sqrt{S}=13~{\rm TeV}$ LHC with $|y_t|<3$ and $|y_J|\leq 4.5$ (a), or $3.0\leq|y_J|\leq 4.5$ (b) . The resummation and renormalization scales are choose as $\mu=\mu_{\rm Res}=\mu_{\rm ren}=H_T$. }
\label{fig:tqt13}
\end{figure}

\noindent {\bf Phenomenology:~}%
Below, we present the numerical result of resummation calculation for the
$t$-channel single top  quark
production at the $\sqrt{S}=13~{\rm TeV}$ LHC with CT14NNLO PDF~\cite{Dulat:2015mca}.
Figure~\ref{fig:tqt13} shows the $q_{\perp}$ distribution from the asymptotic piece (blue dashed line), NLO calculation (red dotted  line), resummation prediction (black solid line) and $Y$-term (orange dot-dashed line) for the top quark  production. Here, the asymptotic piece is the fixed-order expansion of Eq.~(\ref{resumy}) up to the $\alpha_s$ order, and
is expected to agree with the NLO prediction as $q_{\perp} \to 0$.
In the same figure, we also compare to the prediction from the
parton shower event generator PYTHIA 8~\cite{Sjostrand:2007gs} (green solid  line),
which was calculated at the leading order, with CT14LO PDF and $\alpha_s(M_Z)=0.118$ 
at the $Z$-boson mass scale (91.118 GeV). 
For the fixed-order calculation, both the renormalization and factorization scales are fixed at
$H_{T}\equiv\sqrt{m_t^2+P_{J\perp}^2}+P_{J\perp}$.
Similarly, in the resummation calculation, the canonical choice of
the resummation ($\mu_{\rm Res}$) and renormalization ($\mu_{\rm ren}$) scales
is taken to be $H_T$ in this study.
The jet cone size is taken to be $R=0.4$,
using the anti-$k_T$ algorithm, and the Wolfenstein CKM matrix element parameterization is used in our numerical calculation~\cite{Olive:2016xmw}.
We shall compare predictions for two different sets of kinematic cuts, with
$|y_t|\leq 3$ and $P_{J\perp}>30~{\rm GeV}$, and
 $|y_J|\leq 4.5$ in (a), and
$3\leq |y_J|\leq 4.5$ in (b) of Fig.~\ref{fig:tqt13},
respectively.
Some results of the comparison are in order.
Clearly, the asymptotic piece and the fixed-order calculation results agree very well in the small
$q_\perp$ (less than $1~{\rm GeV}$) region.
As a further check, 
we calculated the NLO total cross section predicted by our resummation calculation.
Specifically, we numerically integrated out the $q_{\perp}$ distribution predicted 
 by our resummation calculation from 0 to 1 GeV,
and summed it up with the integration of the perturbative piece (at the $\alpha_s$ order) 
from 1 GeV up to the allowed kinematic region~\cite{Balazs:1997xd}. 
We found that the NLO total cross section predicted 
by our resummation framework and MCFM~\cite{Campbell:2015qma} calculations are in perfect agreement.

\begin{figure}
\includegraphics[width=0.238\textwidth]{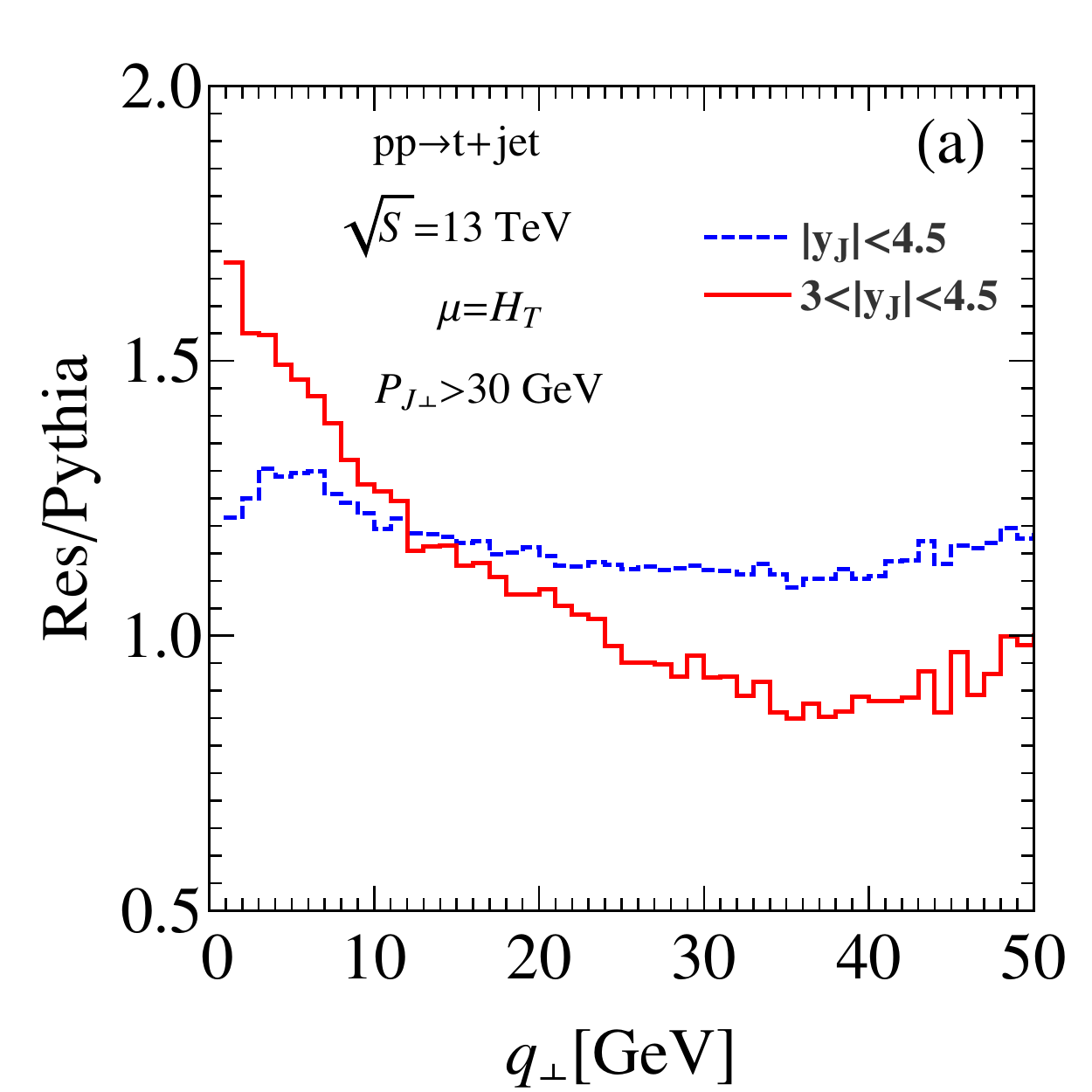}
\includegraphics[width=0.238\textwidth]{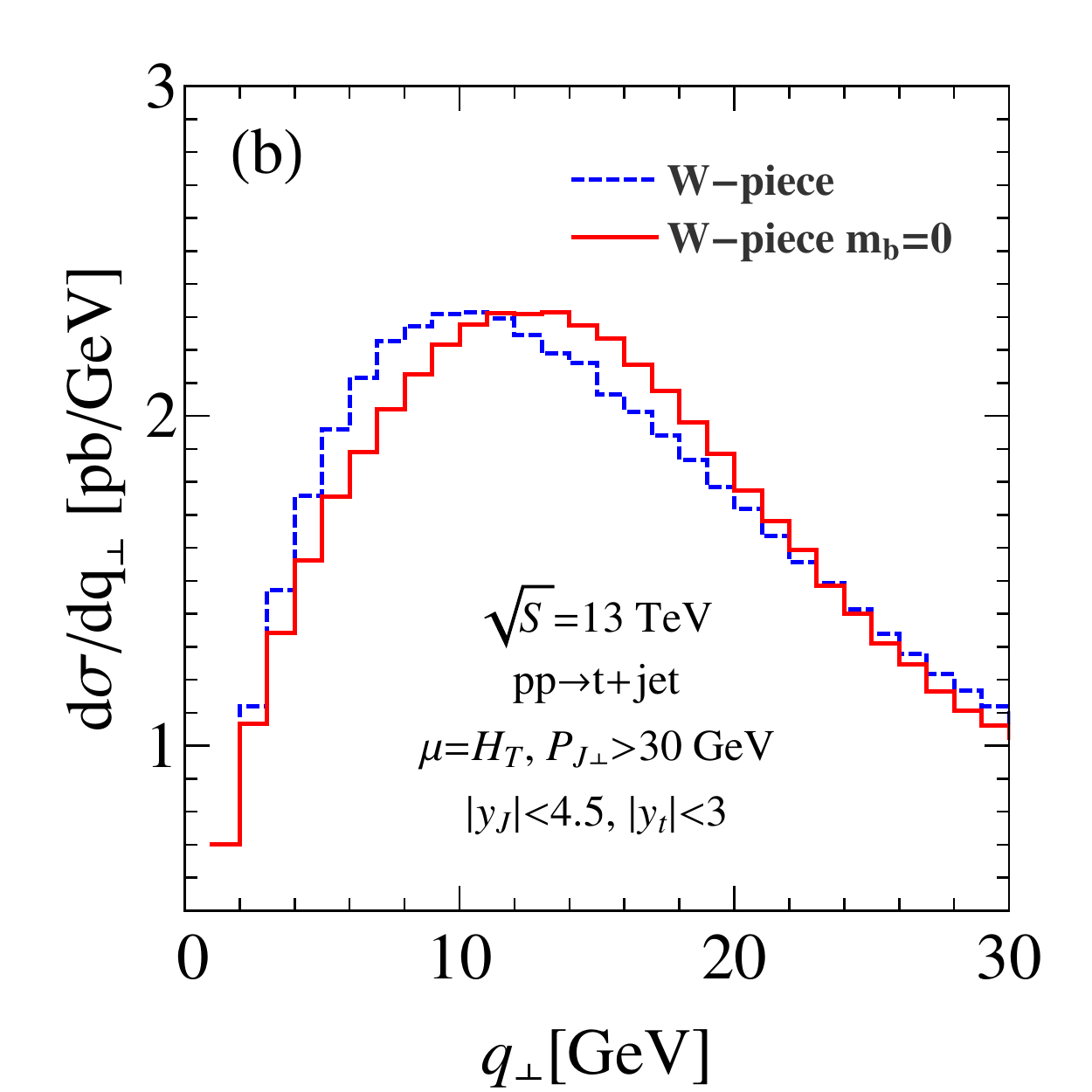}
\caption{(a) The ratio of the resummation and PYTHIA prediction for the $t$-channel single top quark production at the $\sqrt{S}=13~{\rm TeV}$ LHC with $|y_t|<3$, $P_{J\perp}>30~{\rm GeV}$  and $|y_J|\leq 4.5$ (blue dashed line), or $3.0\leq|y_J|\leq 4.5$ (red solid line); (b) The $W$-piece prediction for the single top quark production process with  $m_b=4.75~{\rm GeV}$ (blue dashed line ) and $m_b=0$ (red solid line) at the $\sqrt{S}=13~{\rm TeV}$ LHC with $|y_J|\leq 4.5$, $|y_t|<3$ and $P_{J\perp}>30~{\rm GeV}$. The resummation and renormalization scales are choose as $\mu=\mu_{\rm Res}=\mu_{\rm ren}=H_T$. }
\label{fig:ratio_nb}
\end{figure}

As shown in Fig.~\ref{fig:tqt13}, the NLO prediction is not reliable when the $q_{\perp}$ is small. The resummation calculation predicts a well behavior $q_{\perp}$ distribution in the small $q_{\perp}$ region since the large logarithms have been properly resummed.
In Fig~\ref{fig:ratio_nb}(a), we compare the predictions from our resummation calculation to PYTHIA by taking the ratio of
their $q_{\perp}$ differential distributions shown in Fig.~\ref{fig:tqt13}.
With the jet rapidity $|y_J|\leq 4.5$ (blue dashed line), 
this ratio does not vary strongly with $q_{\perp}$.
Hence, they predict almost the same shape in the $q_{\perp}$ distribution, while
they predict different fiducial total cross sections because PYTHIA prediction includes only
leading order matrix element and is calculated with CT14LO PDFs.
However, if we require the final state jet to be in the forward rapidity region,
with $3\leq |y_J|\leq 4.5$ (red solid line), which is the so-called signal region of single top events,
we find that PYTHIA prediction disagrees with our resummation calculation.
Our resummation calculation predicts a smaller $q_{\perp}$ value when the final state
jet is required to fall into the forward region.
We have checked that the PYTHIA result is not sensitive to the effects from
beam remnants. Furthermore,
the $Y$-term contribution, from NLO, is negligible in this region, cf.
Fig.~\ref{fig:tqt13}(b) (orange dot-dashed line).
Hence, we conclude that their difference most likely comes from the
treatment of multiple soft gluon radiation.

As shown in Eqs.~(\ref{eq:gammas})-(\ref{sud15}),
the effect of multiple gluon radiation, originated from
soft gluons connecting the initial and final state gauge links,
becomes more important when the final state jet is
required to be in the forward region where the
kinematic factor $T\sim \ln\dfrac{-\hat{t}}{\hat{s}}$
becomes large as $|\hat{t}| \to 0$. Consequently,
the $q_\perp$ distribution peaks  at  a smaller value  as  compared to the
case in which the final state jet does not go into the forward region.

Next, we examine the effect of the incoming bottom quark mass
to the $q_{\perp}$ distribution.
As shown in Fig.~\ref{fig:ratio_nb}(b), a finite bottom quark mass,
with $m_b=4.75$ GeV, shifts the peak of the $q_{\perp}$ distribution by about
$3\sim 4~{\rm GeV}$ as compared to massless case.

\begin{figure}
\includegraphics[width=0.238\textwidth]{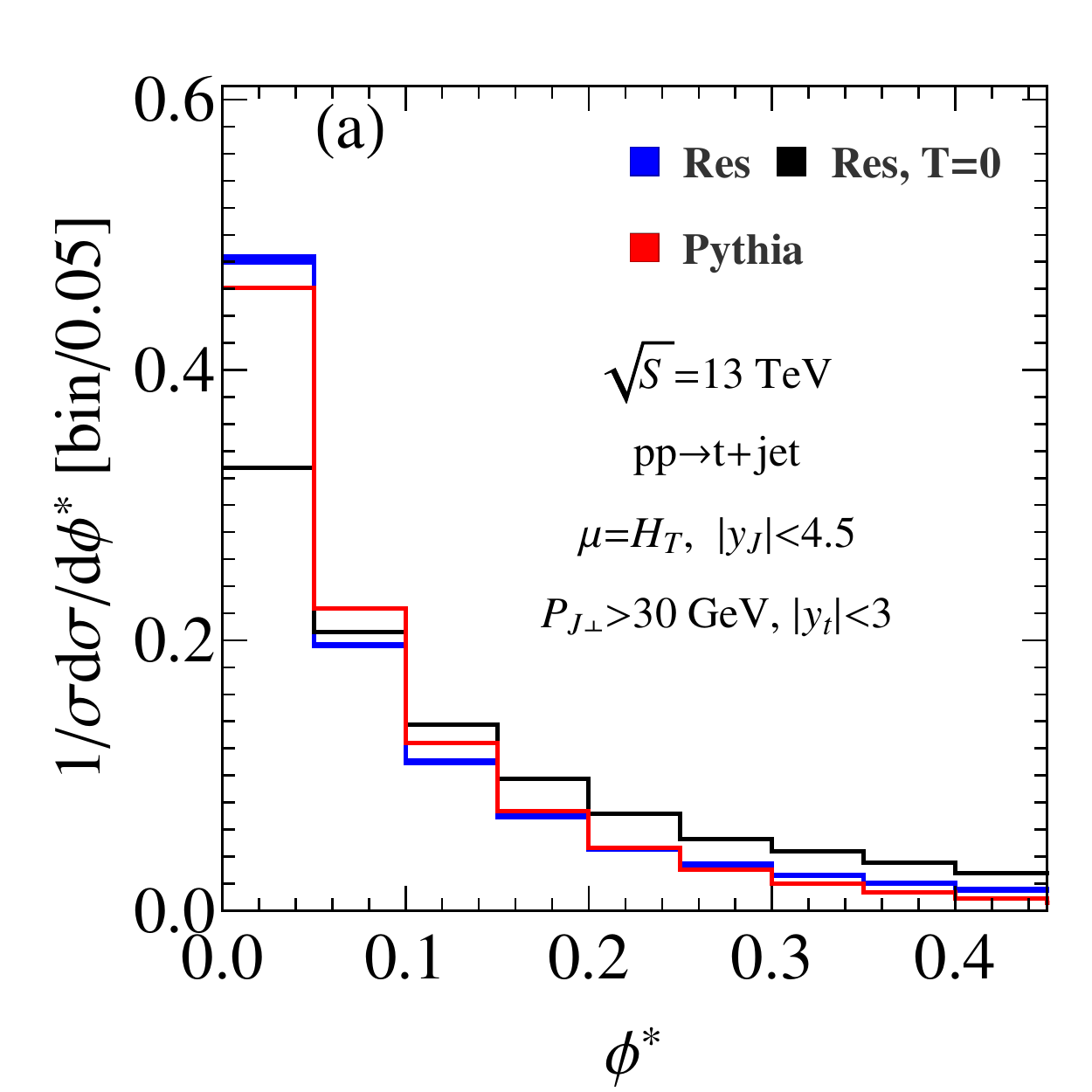}
\includegraphics[width=0.238\textwidth]{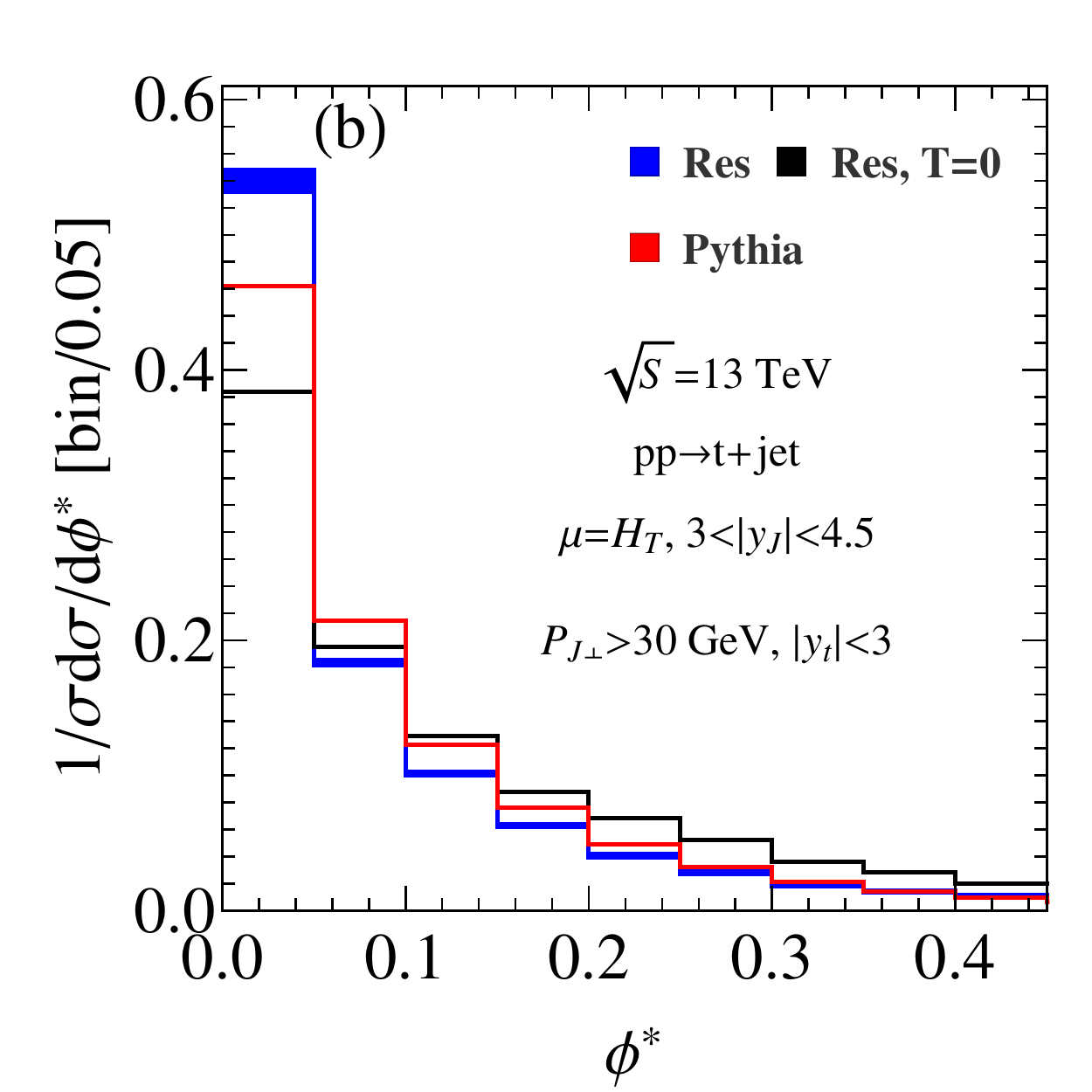}
\caption{The normalized distribution of $\phi^*$ for top quark production at the $\sqrt{S}=13~{\rm TeV}$ LHC with $|y_t|<3$ and $P_{J\perp}>30~{\rm GeV}$. The resummation and renormalization scales are choose as $\mu=\mu_{\rm Res}=\mu_{\rm ren}=H_T$. The blue and black line represents the resummation prediction with and without including the factor  $T$ in Eqs.~(\ref{eq:gammas})-(\ref{sud15}), respectively. The  red lines describe the results from PYTHIA prediction. The blue shaded region represents the scale uncertainties which are varied from $H_T/2$ to $2H_T$.}
\label{fig:phi}
\end{figure}

\begin{table*}[t]
\begin{center}
\caption{The predicted kinematic acceptances for the $\phi^*$ cut-off in the $t$-channel single top quark production at the LHC}
\begin{tabular}{c|c|c|c|c|c|c}
\hline
~~$\phi^*$~~&~~$<0.05$~~&~~$<0.1$~~&~~$<0.15$~~&
~~$<0.2$~~&~~$<0.25$~~&~$<0.3$~\\
\hline
Res~~~~  $|y_J|<4.5$& 48\%& 68 \%& 79\%& 86\%& 91\%& 94\%\\
\hline
PYTHIA $|y_J|<4.5$& 46\% & 68\% & 81\% & 88\% &  93\%&  96\%  \\
\hline
\hline
Res~~~~  $3<|y_J|<4.5$& 54\%& 72\%& 83\%& 89\%& 93\%& 96\%\\
\hline
PYTHIA $3<|y_J|<4.5$& 46\% & 68\% & 80\% & 87\% &  92\%&  96\%  \\
\hline
\end{tabular}
\label{tbl:phicut}
\end{center}
\end{table*}

As discussed above, the coherence effect of gluon radiation in the initial and final states
becomes large when the final state jet falls into more forward (or backward) direction, with
a larger absolute value of pseudorapidity. Furthermore, a different prediction in $q_\perp$
would lead to different prediction in the azimuthal angle between the final state jet and the
top quark moving directions measured in the laboratory frame. Both of them suggest that we
could use the well-known $\phi^*$ distribution, for describing the precision
Drell-Yan pair kinematical distributions~\cite{Banfi:2010cf}, to test the effect of multiple gluon
radiation in the $t$-channel single top quark production.
The advantage of studying the $\phi^*$ distribution is that it only depends on the
moving directions (not energies) of the final state jet and top quark.
Hence, it might provide a more sensitive experimental observable when the
final state jet falls into forward (or backward) direction.
We follow its usual definition and define
\bea
\phi^*=\tan\left(\frac{\pi-\Delta\phi}{2}\right) \sin \theta^*_\eta,
\eea
where $\Delta\phi$ is the azimuthal angle separation in radians
between the jet and top quark. The angle $\theta^*_\eta$ is defined as
\bea
\cos\theta^*_\eta=\tanh\left[\frac{\eta_J-\eta_t}{2}\right],
\eea
where $\eta_J$ and $\eta_t$ are the pseudorapidities of the jet and top quark, respectively.

As shown in Fig.~\ref{fig:phi}, the predictions of PYTHIA and our resumamtion calculation
differ in the small $\phi^*$ region, especially for the final state jet falls into more
forward (or backward) direction (Fig.~\ref{fig:phi}(b)), which can be caused by a large value of $\eta_J-\eta_t$. i.e.,
in the events with large rapidity gap. In such region, the subleading logarithm terms in
the Sudakov factor are important in our resummation calculation. To illustrate this, we also compare to the prediction (shown as black curves in Fig.~\ref{fig:phi})
without the coherence factor $T$ in Eqs.~(\ref{eq:gammas})-(\ref{sud15}). It shows that factor $T$ would change $\phi^*$ distribution significantly.

Since  $\phi^*$ distribution is sensitive to the color structure of the signal, it could also be used to improve the $t$-channel single top quark cross section measurement. In  that case,  a precise theoretical evaluation of the kinematic acceptance is necessary, which is defined as,
\bea
\epsilon\equiv \dfrac{\sigma(\phi^*<\phi^0)}{\sigma}.
\eea
Here, $\sigma(\phi^*<\phi^0)$ is  the cross section after imposing  the kinematic cuts, while $\sigma$ is not.
As shown in Table~\ref{tbl:phicut},  if we require the final state jet to be in the forward rapidity region, with $3\leq|y_J|\leq 4.5$, the  kinematic acceptance with $\phi^*<0.05$ is larger by about 8\% in our resummation calculation than the PYTHIA prediction. For $\phi^*<0.1$, they differ by about 4\%, and our resummation calculation predicts a larger total fiducial cross section. Currently, the ATLAS and CMS Collaborations have measured the $t$-channel single top quark  at the 13 TeV LHC,  the uncertainty is around 10\%~\cite{CMS:2016ayb,Aaboud:2016ymp}. If the $\phi^*$ observable is used to further suppress backgrounds and enhance the signal to backgrounds ratio, the difference found in our resummation and PYTHIA calculations of the fiducial cross section could become important.  This will lead to, for example, different conclusion about the constraints on various $Wtb$ anomalous couplings, induced by New Physics, or the measurement of $V_{tb}$~\cite{Cao:2015doa,Cao:2015qta}.

In summary, we have presented a transverse momentum resummation calculation 
to precisely predict the kinematical distributions of the final state jet and top quark
produced in the $t$-channel single top events at the LHC.
We find that it is important to correctly take into account the color coherence effect, induced by
soft gluons connecting the initial and final states, which 
becomes more significant when the final state jet falls into the more 
forward (or backward) region, where PYTHIA prediction differs the most from our resummation 
calculation. Motivated by this, we propose to apply the experimental observable $\phi^*$, similar the one used in analyzing the precision Drell-Yan data, to perform precision test of the SM in the 
production of the $t$-channel single top events at the LHC.

\begin{acknowledgments}
This work is partially supported by the U.S. Department of Energy,
Office of Science, Office of Nuclear Physics, under contract number
DE-AC02-05CH11231; by the U.S. National
Science Foundation under Grant No. PHY-1719914; and
by the National Natural Science Foundation of China under Grant
Nos. 11275009, 11675002, 11635001 and 11725520.
C.-P. Yuan is also grateful for the support from 
the Wu-Ki Tung endowed chair in particle physics.
\end{acknowledgments}

\bibliographystyle{apsrev}
\bibliography{reference}

\begin{thebibliography}{49}
\expandafter\ifx\csname natexlab\endcsname\relax\def\natexlab#1{#1}\fi
\expandafter\ifx\csname bibnamefont\endcsname\relax
  \def\bibnamefont#1{#1}\fi
\expandafter\ifx\csname bibfnamefont\endcsname\relax
  \def\bibfnamefont#1{#1}\fi
\expandafter\ifx\csname citenamefont\endcsname\relax
  \def\citenamefont#1{#1}\fi
\expandafter\ifx\csname url\endcsname\relax
  \def\url#1{\texttt{#1}}\fi
\expandafter\ifx\csname urlprefix\endcsname\relax\def\urlprefix{URL }\fi
\providecommand{\bibinfo}[2]{#2}
\providecommand{\eprint}[2][]{\url{#2}}

\bibitem[{\citenamefont{Dawson}(1985)}]{Dawson:1984gx}
\bibinfo{author}{\bibfnamefont{S.}~\bibnamefont{Dawson}},
  \bibinfo{journal}{Nucl. Phys.} \textbf{\bibinfo{volume}{B249}},
  \bibinfo{pages}{42} (\bibinfo{year}{1985}).

\bibitem[{\citenamefont{Willenbrock and Dicus}(1986)}]{Willenbrock:1986cr}
\bibinfo{author}{\bibfnamefont{S.~S.~D.} \bibnamefont{Willenbrock}}
  \bibnamefont{and} \bibinfo{author}{\bibfnamefont{D.~A.} \bibnamefont{Dicus}},
  \bibinfo{journal}{Phys. Rev.} \textbf{\bibinfo{volume}{D34}},
  \bibinfo{pages}{155} (\bibinfo{year}{1986}).

\bibitem[{\citenamefont{Dawson and Willenbrock}(1987)}]{Dawson:1986tc}
\bibinfo{author}{\bibfnamefont{S.}~\bibnamefont{Dawson}} \bibnamefont{and}
  \bibinfo{author}{\bibfnamefont{S.~S.~D.} \bibnamefont{Willenbrock}},
  \bibinfo{journal}{Nucl. Phys.} \textbf{\bibinfo{volume}{B284}},
  \bibinfo{pages}{449} (\bibinfo{year}{1987}).

\bibitem[{\citenamefont{Yuan}(1990)}]{Yuan:1989tc}
\bibinfo{author}{\bibfnamefont{C.~P.} \bibnamefont{Yuan}},
  \bibinfo{journal}{Phys. Rev.} \textbf{\bibinfo{volume}{D41}},
  \bibinfo{pages}{42} (\bibinfo{year}{1990}).

\bibitem[{\citenamefont{Berger et~al.}(2017)\citenamefont{Berger, Gao, and
  Zhu}}]{Berger:2017zof}
\bibinfo{author}{\bibfnamefont{E.~L.} \bibnamefont{Berger}},
  \bibinfo{author}{\bibfnamefont{J.}~\bibnamefont{Gao}}, \bibnamefont{and}
  \bibinfo{author}{\bibfnamefont{H.~X.} \bibnamefont{Zhu}}
  (\bibinfo{year}{2017}), \eprint{1708.09405}.

\bibitem[{\citenamefont{Kidonakis}(2006)}]{Kidonakis:2006bu}
\bibinfo{author}{\bibfnamefont{N.}~\bibnamefont{Kidonakis}},
  \bibinfo{journal}{Phys. Rev.} \textbf{\bibinfo{volume}{D74}},
  \bibinfo{pages}{114012} (\bibinfo{year}{2006}), \eprint{hep-ph/0609287}.

\bibitem[{\citenamefont{Kidonakis}(2007)}]{Kidonakis:2007ej}
\bibinfo{author}{\bibfnamefont{N.}~\bibnamefont{Kidonakis}},
  \bibinfo{journal}{Phys. Rev.} \textbf{\bibinfo{volume}{D75}},
  \bibinfo{pages}{071501} (\bibinfo{year}{2007}), \eprint{hep-ph/0701080}.

\bibitem[{\citenamefont{Zhu et~al.}(2011)\citenamefont{Zhu, Li, Wang, and
  Zhang}}]{Zhu:2010mr}
\bibinfo{author}{\bibfnamefont{H.~X.} \bibnamefont{Zhu}},
  \bibinfo{author}{\bibfnamefont{C.~S.} \bibnamefont{Li}},
  \bibinfo{author}{\bibfnamefont{J.}~\bibnamefont{Wang}}, \bibnamefont{and}
  \bibinfo{author}{\bibfnamefont{J.~J.} \bibnamefont{Zhang}},
  \bibinfo{journal}{JHEP} \textbf{\bibinfo{volume}{02}}, \bibinfo{pages}{099}
  (\bibinfo{year}{2011}), \eprint{1006.0681}.

\bibitem[{\citenamefont{Wang et~al.}(2010)\citenamefont{Wang, Li, Zhu, and
  Zhang}}]{Wang:2010ue}
\bibinfo{author}{\bibfnamefont{J.}~\bibnamefont{Wang}},
  \bibinfo{author}{\bibfnamefont{C.~S.} \bibnamefont{Li}},
  \bibinfo{author}{\bibfnamefont{H.~X.} \bibnamefont{Zhu}}, \bibnamefont{and}
  \bibinfo{author}{\bibfnamefont{J.~J.} \bibnamefont{Zhang}}
  (\bibinfo{year}{2010}), \eprint{1010.4509}.

\bibitem[{\citenamefont{Kidonakis}(2011)}]{Kidonakis:2011wy}
\bibinfo{author}{\bibfnamefont{N.}~\bibnamefont{Kidonakis}},
  \bibinfo{journal}{Phys. Rev.} \textbf{\bibinfo{volume}{D83}},
  \bibinfo{pages}{091503} (\bibinfo{year}{2011}), \eprint{1103.2792}.

\bibitem[{\citenamefont{Wang et~al.}(2013)\citenamefont{Wang, Li, and
  Zhu}}]{Wang:2012dc}
\bibinfo{author}{\bibfnamefont{J.}~\bibnamefont{Wang}},
  \bibinfo{author}{\bibfnamefont{C.~S.} \bibnamefont{Li}}, \bibnamefont{and}
  \bibinfo{author}{\bibfnamefont{H.~X.} \bibnamefont{Zhu}},
  \bibinfo{journal}{Phys. Rev.} \textbf{\bibinfo{volume}{D87}},
  \bibinfo{pages}{034030} (\bibinfo{year}{2013}), \eprint{1210.7698}.

\bibitem[{\citenamefont{Collins et~al.}(1985)\citenamefont{Collins, Soper, and
  Sterman}}]{Collins:1984kg}
\bibinfo{author}{\bibfnamefont{J.~C.} \bibnamefont{Collins}},
  \bibinfo{author}{\bibfnamefont{D.~E.} \bibnamefont{Soper}}, \bibnamefont{and}
  \bibinfo{author}{\bibfnamefont{G.~F.} \bibnamefont{Sterman}},
  \bibinfo{journal}{Nucl. Phys.} \textbf{\bibinfo{volume}{B250}},
  \bibinfo{pages}{199} (\bibinfo{year}{1985}).

\bibitem[{\citenamefont{Collins and Soper}(1981)}]{Collins:1981uk}
\bibinfo{author}{\bibfnamefont{J.~C.} \bibnamefont{Collins}} \bibnamefont{and}
  \bibinfo{author}{\bibfnamefont{D.~E.} \bibnamefont{Soper}},
  \bibinfo{journal}{Nucl. Phys.} \textbf{\bibinfo{volume}{B193}},
  \bibinfo{pages}{381} (\bibinfo{year}{1981}), \bibinfo{note}{[Erratum: Nucl.
  Phys.B213,545(1983)]}.

\bibitem[{\citenamefont{Collins and Soper}(1982)}]{Collins:1981va}
\bibinfo{author}{\bibfnamefont{J.~C.} \bibnamefont{Collins}} \bibnamefont{and}
  \bibinfo{author}{\bibfnamefont{D.~E.} \bibnamefont{Soper}},
  \bibinfo{journal}{Nucl. Phys.} \textbf{\bibinfo{volume}{B197}},
  \bibinfo{pages}{446} (\bibinfo{year}{1982}).

\bibitem[{\citenamefont{Zhu et~al.}(2013{\natexlab{a}})\citenamefont{Zhu, Li,
  Li, Shao, and Yang}}]{Zhu:2012ts}
\bibinfo{author}{\bibfnamefont{H.~X.} \bibnamefont{Zhu}},
  \bibinfo{author}{\bibfnamefont{C.~S.} \bibnamefont{Li}},
  \bibinfo{author}{\bibfnamefont{H.~T.} \bibnamefont{Li}},
  \bibinfo{author}{\bibfnamefont{D.~Y.} \bibnamefont{Shao}}, \bibnamefont{and}
  \bibinfo{author}{\bibfnamefont{L.~L.} \bibnamefont{Yang}},
  \bibinfo{journal}{Phys. Rev. Lett.} \textbf{\bibinfo{volume}{110}},
  \bibinfo{pages}{082001} (\bibinfo{year}{2013}{\natexlab{a}}),
  \eprint{1208.5774}.

\bibitem[{\citenamefont{Li et~al.}(2013)\citenamefont{Li, Li, Shao, Yang, and
  Zhu}}]{Li:2013mia}
\bibinfo{author}{\bibfnamefont{H.~T.} \bibnamefont{Li}},
  \bibinfo{author}{\bibfnamefont{C.~S.} \bibnamefont{Li}},
  \bibinfo{author}{\bibfnamefont{D.~Y.} \bibnamefont{Shao}},
  \bibinfo{author}{\bibfnamefont{L.~L.} \bibnamefont{Yang}}, \bibnamefont{and}
  \bibinfo{author}{\bibfnamefont{H.~X.} \bibnamefont{Zhu}},
  \bibinfo{journal}{Phys. Rev.} \textbf{\bibinfo{volume}{D88}},
  \bibinfo{pages}{074004} (\bibinfo{year}{2013}), \eprint{1307.2464}.

\bibitem[{\citenamefont{Zhu et~al.}(2013{\natexlab{b}})\citenamefont{Zhu, Sun,
  and Yuan}}]{Zhu:2013yxa}
\bibinfo{author}{\bibfnamefont{R.}~\bibnamefont{Zhu}},
  \bibinfo{author}{\bibfnamefont{P.}~\bibnamefont{Sun}}, \bibnamefont{and}
  \bibinfo{author}{\bibfnamefont{F.}~\bibnamefont{Yuan}},
  \bibinfo{journal}{Phys. Lett.} \textbf{\bibinfo{volume}{B727}},
  \bibinfo{pages}{474} (\bibinfo{year}{2013}{\natexlab{b}}),
  \eprint{1309.0780}.

\bibitem[{\citenamefont{Sun et~al.}(2015{\natexlab{a}})\citenamefont{Sun, Yuan,
  and Yuan}}]{Sun:2014lna}
\bibinfo{author}{\bibfnamefont{P.}~\bibnamefont{Sun}},
  \bibinfo{author}{\bibfnamefont{C.~P.} \bibnamefont{Yuan}}, \bibnamefont{and}
  \bibinfo{author}{\bibfnamefont{F.}~\bibnamefont{Yuan}},
  \bibinfo{journal}{Phys. Rev. Lett.} \textbf{\bibinfo{volume}{114}},
  \bibinfo{pages}{202001} (\bibinfo{year}{2015}{\natexlab{a}}),
  \eprint{1409.4121}.

\bibitem[{\citenamefont{Sun et~al.}(2014)\citenamefont{Sun, Yuan, and
  Yuan}}]{Sun:2014gfa}
\bibinfo{author}{\bibfnamefont{P.}~\bibnamefont{Sun}},
  \bibinfo{author}{\bibfnamefont{C.~P.} \bibnamefont{Yuan}}, \bibnamefont{and}
  \bibinfo{author}{\bibfnamefont{F.}~\bibnamefont{Yuan}},
  \bibinfo{journal}{Phys. Rev. Lett.} \textbf{\bibinfo{volume}{113}},
  \bibinfo{pages}{232001} (\bibinfo{year}{2014}), \eprint{1405.1105}.

\bibitem[{\citenamefont{Sun et~al.}(2015{\natexlab{b}})\citenamefont{Sun, Yuan,
  and Yuan}}]{Sun:2015doa}
\bibinfo{author}{\bibfnamefont{P.}~\bibnamefont{Sun}},
  \bibinfo{author}{\bibfnamefont{C.~P.} \bibnamefont{Yuan}}, \bibnamefont{and}
  \bibinfo{author}{\bibfnamefont{F.}~\bibnamefont{Yuan}},
  \bibinfo{journal}{Phys. Rev.} \textbf{\bibinfo{volume}{D92}},
  \bibinfo{pages}{094007} (\bibinfo{year}{2015}{\natexlab{b}}),
  \eprint{1506.06170}.

\bibitem[{\citenamefont{Sun et~al.}(2016{\natexlab{a}})\citenamefont{Sun, Yuan,
  and Yuan}}]{Sun:2016mas}
\bibinfo{author}{\bibfnamefont{P.}~\bibnamefont{Sun}},
  \bibinfo{author}{\bibfnamefont{C.~P.} \bibnamefont{Yuan}}, \bibnamefont{and}
  \bibinfo{author}{\bibfnamefont{F.}~\bibnamefont{Yuan}},
  \bibinfo{journal}{Phys. Lett.} \textbf{\bibinfo{volume}{B762}},
  \bibinfo{pages}{47} (\bibinfo{year}{2016}{\natexlab{a}}),
  \eprint{1605.00063}.

\bibitem[{\citenamefont{Sun et~al.}(2016{\natexlab{b}})\citenamefont{Sun,
  Isaacson, Yuan, and Yuan}}]{Sun:2016kkh}
\bibinfo{author}{\bibfnamefont{P.}~\bibnamefont{Sun}},
  \bibinfo{author}{\bibfnamefont{J.}~\bibnamefont{Isaacson}},
  \bibinfo{author}{\bibfnamefont{C.~P.} \bibnamefont{Yuan}}, \bibnamefont{and}
  \bibinfo{author}{\bibfnamefont{F.}~\bibnamefont{Yuan}}
  (\bibinfo{year}{2016}{\natexlab{b}}), \eprint{1602.08133}.

\bibitem[{\citenamefont{Xiao and Yuan}(2018)}]{Xiao:2018esv}
\bibinfo{author}{\bibfnamefont{B.-W.} \bibnamefont{Xiao}} \bibnamefont{and}
  \bibinfo{author}{\bibfnamefont{F.}~\bibnamefont{Yuan}}
  (\bibinfo{year}{2018}), \eprint{1801.05478}.

\bibitem[{\citenamefont{Landry et~al.}(2001)\citenamefont{Landry, Brock,
  Ladinsky, and Yuan}}]{Landry:1999an}
\bibinfo{author}{\bibfnamefont{F.}~\bibnamefont{Landry}},
  \bibinfo{author}{\bibfnamefont{R.}~\bibnamefont{Brock}},
  \bibinfo{author}{\bibfnamefont{G.}~\bibnamefont{Ladinsky}}, \bibnamefont{and}
  \bibinfo{author}{\bibfnamefont{C.~P.} \bibnamefont{Yuan}},
  \bibinfo{journal}{Phys. Rev.} \textbf{\bibinfo{volume}{D63}},
  \bibinfo{pages}{013004} (\bibinfo{year}{2001}), \eprint{hep-ph/9905391}.

\bibitem[{\citenamefont{Landry et~al.}(2003)\citenamefont{Landry, Brock,
  Nadolsky, and Yuan}}]{Landry:2002ix}
\bibinfo{author}{\bibfnamefont{F.}~\bibnamefont{Landry}},
  \bibinfo{author}{\bibfnamefont{R.}~\bibnamefont{Brock}},
  \bibinfo{author}{\bibfnamefont{P.~M.} \bibnamefont{Nadolsky}},
  \bibnamefont{and} \bibinfo{author}{\bibfnamefont{C.~P.} \bibnamefont{Yuan}},
  \bibinfo{journal}{Phys. Rev.} \textbf{\bibinfo{volume}{D67}},
  \bibinfo{pages}{073016} (\bibinfo{year}{2003}), \eprint{hep-ph/0212159}.

\bibitem[{\citenamefont{Sun et~al.}(2013)\citenamefont{Sun, Yuan, and
  Yuan}}]{Sun:2012vc}
\bibinfo{author}{\bibfnamefont{P.}~\bibnamefont{Sun}},
  \bibinfo{author}{\bibfnamefont{C.~P.} \bibnamefont{Yuan}}, \bibnamefont{and}
  \bibinfo{author}{\bibfnamefont{F.}~\bibnamefont{Yuan}},
  \bibinfo{journal}{Phys. Rev.} \textbf{\bibinfo{volume}{D88}},
  \bibinfo{pages}{054008} (\bibinfo{year}{2013}), \eprint{1210.3432}.

\bibitem[{\citenamefont{Dulat et~al.}(2016)\citenamefont{Dulat, Hou, Gao,
  Guzzi, Huston, Nadolsky, Pumplin, Schmidt, Stump, and Yuan}}]{Dulat:2015mca}
\bibinfo{author}{\bibfnamefont{S.}~\bibnamefont{Dulat}},
  \bibinfo{author}{\bibfnamefont{T.-J.} \bibnamefont{Hou}},
  \bibinfo{author}{\bibfnamefont{J.}~\bibnamefont{Gao}},
  \bibinfo{author}{\bibfnamefont{M.}~\bibnamefont{Guzzi}},
  \bibinfo{author}{\bibfnamefont{J.}~\bibnamefont{Huston}},
  \bibinfo{author}{\bibfnamefont{P.}~\bibnamefont{Nadolsky}},
  \bibinfo{author}{\bibfnamefont{J.}~\bibnamefont{Pumplin}},
  \bibinfo{author}{\bibfnamefont{C.}~\bibnamefont{Schmidt}},
  \bibinfo{author}{\bibfnamefont{D.}~\bibnamefont{Stump}}, \bibnamefont{and}
  \bibinfo{author}{\bibfnamefont{C.~P.} \bibnamefont{Yuan}},
  \bibinfo{journal}{Phys. Rev.} \textbf{\bibinfo{volume}{D93}},
  \bibinfo{pages}{033006} (\bibinfo{year}{2016}), \eprint{1506.07443}.

\bibitem[{\citenamefont{Aivazis
  et~al.}(1994{\natexlab{a}})\citenamefont{Aivazis, Olness, and
  Tung}}]{Aivazis:1993kh}
\bibinfo{author}{\bibfnamefont{M.~A.~G.} \bibnamefont{Aivazis}},
  \bibinfo{author}{\bibfnamefont{F.~I.} \bibnamefont{Olness}},
  \bibnamefont{and} \bibinfo{author}{\bibfnamefont{W.-K.} \bibnamefont{Tung}},
  \bibinfo{journal}{Phys. Rev.} \textbf{\bibinfo{volume}{D50}},
  \bibinfo{pages}{3085} (\bibinfo{year}{1994}{\natexlab{a}}),
  \eprint{hep-ph/9312318}.

\bibitem[{\citenamefont{Aivazis
  et~al.}(1994{\natexlab{b}})\citenamefont{Aivazis, Collins, Olness, and
  Tung}}]{Aivazis:1993pi}
\bibinfo{author}{\bibfnamefont{M.~A.~G.} \bibnamefont{Aivazis}},
  \bibinfo{author}{\bibfnamefont{J.~C.} \bibnamefont{Collins}},
  \bibinfo{author}{\bibfnamefont{F.~I.} \bibnamefont{Olness}},
  \bibnamefont{and} \bibinfo{author}{\bibfnamefont{W.-K.} \bibnamefont{Tung}},
  \bibinfo{journal}{Phys. Rev.} \textbf{\bibinfo{volume}{D50}},
  \bibinfo{pages}{3102} (\bibinfo{year}{1994}{\natexlab{b}}),
  \eprint{hep-ph/9312319}.

\bibitem[{\citenamefont{Collins}(1998)}]{Collins:1998rz}
\bibinfo{author}{\bibfnamefont{J.~C.} \bibnamefont{Collins}},
  \bibinfo{journal}{Phys. Rev.} \textbf{\bibinfo{volume}{D58}},
  \bibinfo{pages}{094002} (\bibinfo{year}{1998}), \eprint{hep-ph/9806259}.

\bibitem[{\citenamefont{Kramer et~al.}(2000)\citenamefont{Kramer, Olness, and
  Soper}}]{Kramer:2000hn}
\bibinfo{author}{\bibfnamefont{M.}~\bibnamefont{Kramer}},
  \bibinfo{author}{\bibfnamefont{F.~I.} \bibnamefont{Olness}},
  \bibnamefont{and} \bibinfo{author}{\bibfnamefont{D.~E.} \bibnamefont{Soper}},
  \bibinfo{journal}{Phys. Rev.} \textbf{\bibinfo{volume}{D62}},
  \bibinfo{pages}{096007} (\bibinfo{year}{2000}), \eprint{hep-ph/0003035}.

\bibitem[{\citenamefont{Nadolsky et~al.}(2003)\citenamefont{Nadolsky,
  Kidonakis, Olness, and Yuan}}]{Nadolsky:2002jr}
\bibinfo{author}{\bibfnamefont{P.~M.} \bibnamefont{Nadolsky}},
  \bibinfo{author}{\bibfnamefont{N.}~\bibnamefont{Kidonakis}},
  \bibinfo{author}{\bibfnamefont{F.~I.} \bibnamefont{Olness}},
  \bibnamefont{and} \bibinfo{author}{\bibfnamefont{C.~P.} \bibnamefont{Yuan}},
  \bibinfo{journal}{Phys. Rev.} \textbf{\bibinfo{volume}{D67}},
  \bibinfo{pages}{074015} (\bibinfo{year}{2003}), \eprint{hep-ph/0210082}.

\bibitem[{\citenamefont{Belyaev et~al.}(2006)\citenamefont{Belyaev, Nadolsky,
  and Yuan}}]{Belyaev:2005bs}
\bibinfo{author}{\bibfnamefont{A.}~\bibnamefont{Belyaev}},
  \bibinfo{author}{\bibfnamefont{P.~M.} \bibnamefont{Nadolsky}},
  \bibnamefont{and} \bibinfo{author}{\bibfnamefont{C.~P.} \bibnamefont{Yuan}},
  \bibinfo{journal}{JHEP} \textbf{\bibinfo{volume}{04}}, \bibinfo{pages}{004}
  (\bibinfo{year}{2006}), \eprint{hep-ph/0509100}.

\bibitem[{\citenamefont{Berge et~al.}(2006)\citenamefont{Berge, Nadolsky, and
  Olness}}]{Berge:2005rv}
\bibinfo{author}{\bibfnamefont{S.}~\bibnamefont{Berge}},
  \bibinfo{author}{\bibfnamefont{P.~M.} \bibnamefont{Nadolsky}},
  \bibnamefont{and} \bibinfo{author}{\bibfnamefont{F.~I.}
  \bibnamefont{Olness}}, \bibinfo{journal}{Phys. Rev.}
  \textbf{\bibinfo{volume}{D73}}, \bibinfo{pages}{013002}
  (\bibinfo{year}{2006}), \eprint{hep-ph/0509023}.

\bibitem[{\citenamefont{Mukherjee and Vogelsang}(2012)}]{Mukherjee:2012uz}
\bibinfo{author}{\bibfnamefont{A.}~\bibnamefont{Mukherjee}} \bibnamefont{and}
  \bibinfo{author}{\bibfnamefont{W.}~\bibnamefont{Vogelsang}},
  \bibinfo{journal}{Phys. Rev.} \textbf{\bibinfo{volume}{D86}},
  \bibinfo{pages}{094009} (\bibinfo{year}{2012}), \eprint{1209.1785}.

\bibitem[{\citenamefont{Dasgupta and Salam}(2001)}]{Dasgupta:2001sh}
\bibinfo{author}{\bibfnamefont{M.}~\bibnamefont{Dasgupta}} \bibnamefont{and}
  \bibinfo{author}{\bibfnamefont{G.~P.} \bibnamefont{Salam}},
  \bibinfo{journal}{Phys. Lett.} \textbf{\bibinfo{volume}{B512}},
  \bibinfo{pages}{323} (\bibinfo{year}{2001}), \eprint{hep-ph/0104277}.

\bibitem[{\citenamefont{Dasgupta and Salam}(2002)}]{Dasgupta:2002bw}
\bibinfo{author}{\bibfnamefont{M.}~\bibnamefont{Dasgupta}} \bibnamefont{and}
  \bibinfo{author}{\bibfnamefont{G.~P.} \bibnamefont{Salam}},
  \bibinfo{journal}{JHEP} \textbf{\bibinfo{volume}{03}}, \bibinfo{pages}{017}
  (\bibinfo{year}{2002}), \eprint{hep-ph/0203009}.

\bibitem[{\citenamefont{Banfi and Dasgupta}(2004)}]{Banfi:2003jj}
\bibinfo{author}{\bibfnamefont{A.}~\bibnamefont{Banfi}} \bibnamefont{and}
  \bibinfo{author}{\bibfnamefont{M.}~\bibnamefont{Dasgupta}},
  \bibinfo{journal}{JHEP} \textbf{\bibinfo{volume}{01}}, \bibinfo{pages}{027}
  (\bibinfo{year}{2004}), \eprint{hep-ph/0312108}.

\bibitem[{\citenamefont{Forshaw et~al.}(2006)\citenamefont{Forshaw, Kyrieleis,
  and Seymour}}]{Forshaw:2006fk}
\bibinfo{author}{\bibfnamefont{J.~R.} \bibnamefont{Forshaw}},
  \bibinfo{author}{\bibfnamefont{A.}~\bibnamefont{Kyrieleis}},
  \bibnamefont{and} \bibinfo{author}{\bibfnamefont{M.~H.}
  \bibnamefont{Seymour}}, \bibinfo{journal}{JHEP}
  \textbf{\bibinfo{volume}{08}}, \bibinfo{pages}{059} (\bibinfo{year}{2006}),
  \eprint{hep-ph/0604094}.

\bibitem[{\citenamefont{Sun et~al.}(2018)\citenamefont{Sun, Yuan, and
  Yuan}}]{Sun:temp}
\bibinfo{author}{\bibfnamefont{P.}~\bibnamefont{Sun}},
  \bibinfo{author}{\bibfnamefont{C.~P.} \bibnamefont{Yuan}}, \bibnamefont{and}
  \bibinfo{author}{\bibfnamefont{F.}~\bibnamefont{Yuan}}
  (\bibinfo{year}{2018}), \eprint{in preparation}.

\bibitem[{\citenamefont{Sjostrand et~al.}(2008)\citenamefont{Sjostrand, Mrenna,
  and Skands}}]{Sjostrand:2007gs}
\bibinfo{author}{\bibfnamefont{T.}~\bibnamefont{Sjostrand}},
  \bibinfo{author}{\bibfnamefont{S.}~\bibnamefont{Mrenna}}, \bibnamefont{and}
  \bibinfo{author}{\bibfnamefont{P.~Z.} \bibnamefont{Skands}},
  \bibinfo{journal}{Comput. Phys. Commun.} \textbf{\bibinfo{volume}{178}},
  \bibinfo{pages}{852} (\bibinfo{year}{2008}), \eprint{0710.3820}.

\bibitem[{\citenamefont{Patrignani et~al.}(2016)}]{Olive:2016xmw}
\bibinfo{author}{\bibfnamefont{C.}~\bibnamefont{Patrignani}}
  \bibnamefont{et~al.} (\bibinfo{collaboration}{Particle Data Group}),
  \bibinfo{journal}{Chin. Phys.} \textbf{\bibinfo{volume}{C40}},
  \bibinfo{pages}{100001} (\bibinfo{year}{2016}).

\bibitem[{\citenamefont{Balazs and Yuan}(1997)}]{Balazs:1997xd}
\bibinfo{author}{\bibfnamefont{C.}~\bibnamefont{Balazs}} \bibnamefont{and}
  \bibinfo{author}{\bibfnamefont{C.~P.} \bibnamefont{Yuan}},
  \bibinfo{journal}{Phys. Rev.} \textbf{\bibinfo{volume}{D56}},
  \bibinfo{pages}{5558} (\bibinfo{year}{1997}), \eprint{hep-ph/9704258}.

\bibitem[{\citenamefont{Campbell et~al.}(2015)\citenamefont{Campbell, Ellis,
  and Giele}}]{Campbell:2015qma}
\bibinfo{author}{\bibfnamefont{J.~M.} \bibnamefont{Campbell}},
  \bibinfo{author}{\bibfnamefont{R.~K.} \bibnamefont{Ellis}}, \bibnamefont{and}
  \bibinfo{author}{\bibfnamefont{W.~T.} \bibnamefont{Giele}},
  \bibinfo{journal}{Eur. Phys. J.} \textbf{\bibinfo{volume}{C75}},
  \bibinfo{pages}{246} (\bibinfo{year}{2015}), \eprint{1503.06182}.

\bibitem[{\citenamefont{Banfi et~al.}(2011)\citenamefont{Banfi, Redford,
  Vesterinen, Waller, and Wyatt}}]{Banfi:2010cf}
\bibinfo{author}{\bibfnamefont{A.}~\bibnamefont{Banfi}},
  \bibinfo{author}{\bibfnamefont{S.}~\bibnamefont{Redford}},
  \bibinfo{author}{\bibfnamefont{M.}~\bibnamefont{Vesterinen}},
  \bibinfo{author}{\bibfnamefont{P.}~\bibnamefont{Waller}}, \bibnamefont{and}
  \bibinfo{author}{\bibfnamefont{T.~R.} \bibnamefont{Wyatt}},
  \bibinfo{journal}{Eur. Phys. J.} \textbf{\bibinfo{volume}{C71}},
  \bibinfo{pages}{1600} (\bibinfo{year}{2011}), \eprint{1009.1580}.

\bibitem[{\citenamefont{Collaboration}(2016)}]{CMS:2016ayb}
\bibinfo{author}{\bibfnamefont{C.}~\bibnamefont{Collaboration}}
  (\bibinfo{collaboration}{CMS}) (\bibinfo{year}{2016}).

\bibitem[{\citenamefont{Aaboud et~al.}(2017)}]{Aaboud:2016ymp}
\bibinfo{author}{\bibfnamefont{M.}~\bibnamefont{Aaboud}} \bibnamefont{et~al.}
  (\bibinfo{collaboration}{ATLAS}), \bibinfo{journal}{JHEP}
  \textbf{\bibinfo{volume}{04}}, \bibinfo{pages}{086} (\bibinfo{year}{2017}),
  \eprint{1609.03920}.

\bibitem[{\citenamefont{Cao et~al.}(2017)\citenamefont{Cao, Yan, Yu, and
  Zhang}}]{Cao:2015doa}
\bibinfo{author}{\bibfnamefont{Q.-H.} \bibnamefont{Cao}},
  \bibinfo{author}{\bibfnamefont{B.}~\bibnamefont{Yan}},
  \bibinfo{author}{\bibfnamefont{J.-H.} \bibnamefont{Yu}}, \bibnamefont{and}
  \bibinfo{author}{\bibfnamefont{C.}~\bibnamefont{Zhang}},
  \bibinfo{journal}{Chin. Phys.} \textbf{\bibinfo{volume}{C41}},
  \bibinfo{pages}{063101} (\bibinfo{year}{2017}), \eprint{1504.03785}.

\bibitem[{\citenamefont{Cao and Yan}(2015)}]{Cao:2015qta}
\bibinfo{author}{\bibfnamefont{Q.-H.} \bibnamefont{Cao}} \bibnamefont{and}
  \bibinfo{author}{\bibfnamefont{B.}~\bibnamefont{Yan}},
  \bibinfo{journal}{Phys. Rev.} \textbf{\bibinfo{volume}{D92}},
  \bibinfo{pages}{094018} (\bibinfo{year}{2015}), \eprint{1507.06204}.

\end{thebibliography}

\end{document}